\documentclass[aps,prl,twocolumn,superscriptaddress]{revtex4}
\usepackage{epsfig}
\usepackage{graphicx}

\begin{document}
\title{A unified treatment of current-induced instabilities on Si surfaces}                                    
\author{Tong Zhao}
%\email[]{}
%\homepage[]{Your web page}
%\thanks{}
                                                                                                                            
\affiliation{Institute for Physical Science and Technology,
and Department of Chemistry and Biochemistry, University
of Maryland, College Park, MD 20742}
\author{John D. Weeks}
\affiliation{Institute for Physical Science and Technology,
and Department of Chemistry and Biochemistry, University
of Maryland, College Park, MD 20742}
\author{Daniel Kandel}
%\email[]{}
%\homepage[]{Your web page}
%\thanks{}
                                                                                                                            
\affiliation{Department of Physics of Complex System, Weizmann
Institute of Science, Rehovot 76100, Israel}
\date{\today }
 
\begin{abstract}
We introduce a simple two-region model 
where the diffusion constant in a small region around each
step on a vicinal surface can differ from that found on the terraces.
Steady state results for this model provide a physically
suggestive mapping onto kinetic coefficients in the 
conventional sharp-step model, with a negative coefficient
arising from faster diffusion in the step region. A linear stability
analysis of the resulting sharp-step model provides a unified
and simple interpretation of many experimental results
for current-induced step bunching and wandering instabilities
on both Si(111) and Si(001) surfaces.
\end{abstract} 
 
\maketitle

Si surfaces heated with a direct electric current undergo striking
morphological changes induced by step bunching and step wandering instabilities. 
\cite{Latyshev89,yagireview}
These phenomena are great interest not only because
of possible applications for directed growth and nanofabrication, but also
as physical examples of pattern formation on large ($\mu $m) length scales in
a system driven far from equilibrium by a weak and externally controllable field.

We present here a simple model that describes the interplay between the
externally driven diffusion of adatoms by the electric field, and the intrinsic
modulations and anisotropy of the diffusion pathways and the attachment
kinetics on terraces and steps. It provides a new interpretation of the
kinetic coefficients used in traditional step models and shows that negative
kinetic coefficients can arise from faster diffusion in the region around a
step. Many features of the instabilities seen on Si surfaces can be
understood from this perspective.

Step bunching is seen on vicinal Si(111) surfaces when the current is properly
directed normal to the steps.\cite{Latyshev89,yagireview}
The uniform step train is initially stable when the current flows in
the opposite direction. This instability has a mysterious dependence on
temperature, with three temperature ranges between about $850^{\circ }$C and 
$1300^{\circ }$C where the stable and unstable directions are reversed.
Moreover, in temperature range {\rm II }(about $1050^{\circ }$C
to $1150^{\circ }$C) after heating for several hours with a current in the nominally
stable (step-down) direction, the steps undergo a novel {\em wandering
instability} with finite wavelength in-phase sinusoidal undulations in their
positions.\cite{yagireview,Si111wanexp_Degawa3}

Current-induced step bunching also occurs on Si(001) miscut along
$\langle 110\rangle $, but with some notable differences presumably arising
from the alternating (1x2) and (2x1) dimer reconstructions on
adjacent equilibrium terraces. At low temperatures step bunching is found
for current normal to the steps in {\em both }directions, but involving
paired double height steps.
\cite{yagireview,Si001bchexp_Doi,Si001bchexp_Latyshev,Si001bchexp_Nielsen}

The behavior in the lowest temperature range
I of Si(111) is well described by a continuum diffusion model with
nonequilibrium boundary conditions at sharp step edges \cite{Si111bchthy_Liu}
if one assumes that adatoms acquire a small positive effective charge $e^{*}$
and undergo biased diffusion \cite{stoyanov} from a field-dependent
force ${\bf F}= e^{*}{\bf E}$. However, it is not clear how to modify this
picture to account
for the reversals in the stable current direction at higher temperatures.
Experiments \cite{yagireview} have ruled out the simplest explanation, a
change of sign of the effective charge \cite{Effchar_Kandel}, and it
seems likely that different boundary conditions are needed to describe the
wandering instability. \cite{otherattempts}

To gain some insight we introduce here a simple model that can give a more
detailed description of processes occurring in the region around a step. The
characteristic surface reconstructions seen on semiconductor surfaces strongly
affect surface diffusion rates and pathways. We expect a different local
reconstruction of bonds in the vicinity of a step. This suggests it could be
profitable to view a step ``dressed'' by its local reconstruction as
defining a {\em region} of finite width $s$ (of a few atomic spacings $a$)
where adatoms undergo effective diffusion with a diffusion constant $D_{s}$
that can differ from $D_{t}$, the value found elsewhere on the terraces.
\cite{phasefield} For Si(111) we can take $D_{s}$ and $D_{t}$ as
isotropic and assume the step region has a fixed width $s$ at a given
temperature.

Thus a uniform vicinal Si(111) surface can
be viewed as an array of two-region units, made up of the $n$th
step region of width $s$ and its neighboring lower terrace region, with
width $l_{t}\gg s$. We assume that the straight steps extend along the $y$\
direction and the step index increases in the positive $x$\ (step down)
direction, with $x$ measured from the center of the step region. See Fig.
(1).

The biased diffusion flux of adatoms with density $c$ takes the form: ${\bf J%
}_{\alpha }=-D_{\alpha }{\bf \nabla }c_{\alpha }+D_{\alpha }{\bf f}c_{\alpha
}$, where $\alpha =(t,s)$ indicates the terrace or step regions and ${\bf f}%
\equiv {\bf F/}k_{B}T$. (We neglect evaporation and assume a constant
positive effective charge.) We first consider the steady-state solutions
that arise when the electric field is directed normal to the steps and let $%
f\equiv {\bf f}\cdot \hat{x}.$ We can ignore the small effects of step
motion on the steady state adatom density field and determine $c$ by simply
requiring ${\bf \nabla }\cdot {\bf J}_{\alpha }=0$ in each region, along
with continuity of $c$ and ${\bf J}$ at the fixed boundary at $x=s/2$
between the step region and the lower terrace region. In almost all cases of
physical interest, the field is sufficiently weak that $fl_{t}$ and $fs$ are
much less than one, and the steady state profiles are piecewise linear.

In particular the steady state terrace density is
\begin{equation}
\begin{array}{ll}
c_{t}^{0}(x) & =c_{eq}^{0}\left[ 1-%
%TCIMACRO{\dfrac{(l/2-x)(R-1)sf}{l+(R-1)s} }
%BeginExpansion
{\displaystyle {(l/2-x)(R-1)sf \over l+(R-1)s}}%
%EndExpansion
\right] \;\;\;
\end{array}
\label{linterrace}
\end{equation}
for $s/2\leq x\leq l-s/2$, with a similar linear expression for $%
c_{s}^{0}(x).$ The (constant) adatom flux is given by 
\begin{equation}
J_{0}(l)=\frac{D_{t}c_{eq}^{0}f\,l}{l+(R-1)s}.  \label{steadystateflux}
\end{equation}
Here $l\equiv l_{t}+s$ is the distance between the centers of adjacent step
regions, $R\equiv D_{t}/D_{s}$ is a key dimensionless parameter that
describes the relative diffusion rates of adatoms in the terrace and step
regions, and $c_{eq}^{0}$ is the average equilibrium density when $f=0.$
This is also the density at the center of the step (and terrace) region, so
one can view the step region as comprised of a classical local equilibrium
sharp step at $x=0$ surrounded by a symmetric local region of width $s$ with
a different diffusion constant, as illustrated in Fig.\ (1).

%*********
\begin{figure}[tbp]
\includegraphics[width=76mm,height=57mm]{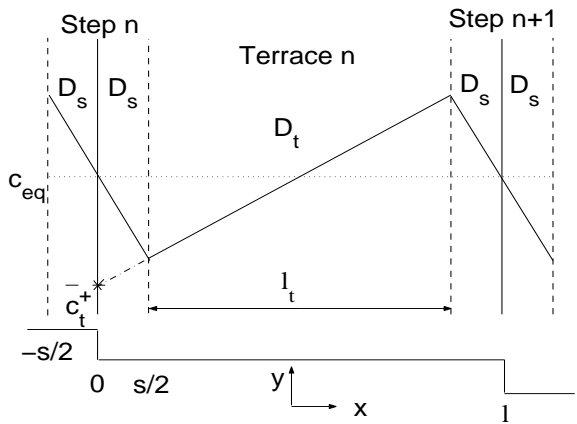}
\caption{Geometry and density profile of the two-region model. The sloping
solid lines in the upper part of the figure give the steady state density
profile as a function of distance $x$ from the center of the left step region, 
denoted by the vertical dashed lines. Shown is a highly exaggerated profile
for a downhill force and slower diffusion in the step region, yielding step
bunching in range I in Si(111). Also illustrated with the dashed-dot line is
the extrapolation of the terrace profile to the center of the step region, 
thus determining the parameter $\bar{c}_{t}^{+}$ in Eq.\ (\ref{kineticbc}). 
The lower part of the figure gives a side view of sharp equilibrium steps
and their associated step regions.}
\end{figure}

Equation (\ref{linterrace}) shows that the slope of the steady-state density
profile is proportional to $(R-1)f.$ Thus, there are four general types of
profiles, arising from a downhill force ($f>0$) or uphill force ($f<0$)
combined with faster diffusion in the terrace region ($R>1$) or in the step
region ($R<1$). We will use these results to make a precise connection between
the physically suggestive parameters of the two-region model and appropriate
boundary conditions in an equivalent symmetric sharp-step model giving the
same steady state terrace profiles. Both bunching and wandering
instabilities can then be readily described using this standard framework.

The general continuum boundary condition in the sharp-step model assumes
small deviations from local equilibrium and introduces linear {\em kinetic
coefficients} $k_{\pm }$ to relate $\bar{c}_{t}^{+}$ (or $\bar{c}_{t}^{-}$),
the limiting lower (or upper) terrace adatom density at the step edge, to
the associated terrace adatom flux into the step: 
\begin{equation}
\mp \hat{n}\cdot {\bf J}_{t}^{\pm }=k_{\pm }(\bar{c}_{t}^{\pm }-c_{eq}).
\label{kineticbc}
\end{equation}
Here $c_{eq}=c_{eq}^{0}[1+\Gamma \kappa ]$ with $\Gamma $ the capillary
length (proportional to the step stiffness) and $\kappa $
is the local step curvature. \cite{Si111bchthy_Liu}
For Si(111) experiments are consistent with a symmetric model where
$k_{+}=k_{-}=k$.

In the usual quasi-stationary approximation, the diffusion field with
boundary conditions given by Eq.\ (\ref{kineticbc}) is first calculated for
fixed step positions, and then the step velocity is determined from the net
local flux and mass conservation: 
\begin{equation}
v_{n}\Delta c=\hat{n}\cdot [{\bf J}_{t}^{-}-{\bf J}_{t}^{+}]-\partial _{\tau
}J_{s}.  \label{normalvelocity}
\end{equation}
Here $\Delta c=\Omega ^{-1}\simeq a^{-2}$ is the change in areal density
when an atom joins the solid, $J_{s}=-D_{s}\partial _{\tau }c_{s}+D_{s}(\hat{%
\tau}\cdot {\bf f})c_{s}$ denotes a tangential or periphery
diffusion flux along the step edge (the sharp-step analogue of parallel
diffusion in the step region of the two-region model), and $c_{s}\simeq
c_{eq}s$ gives the effective number of ledge atoms per unit step length with
diffusion controlled by $D_{s}$ rather than $D_{t}.$

It is natural to identify the terrace width in the appropriate sharp-step
model with $l=l_{t}+s$ and to relate the limiting terrace density $\bar{c}%
_{t}^{+}$ in Eq.\ (\ref{kineticbc}) to $c_{t}^{0}(0)$, the {\em extrapolation}
of the two-region terrace profile in Eq.\ (\ref{linterrace}) to the center of
the step region, as shown in Fig.\ (1). Relating parameters in discrete and
continuum models by extrapolation is well known in other interface
applications \cite{extrapolation}.

Using these results and the flux from Eq.\ (\ref{steadystateflux}) in Eq.\ (%
\ref{kineticbc}), we find to lowest order in $f$ our basic result: 
\begin{equation}
\frac{D_{t}}{k}\equiv d=\frac{1}{2}(R-1)s.  \label{indentifyk}
\end{equation}
This equation relates the fundamental parameters $R$ and $s$ of the simplest
two-region model to the kinetic coefficient $k$ in an equivalent sharp-step
model. \cite{our1dmodel}

Thus a positive kinetic coefficient $k$ can arise from {\em slower
diffusion in the step region} ($R>1$), in accord with
the usual picture of an attachment
barrier in range I. Indeed the extrapolated profile in Fig.\ (1)
corresponds exactly to the linear steady state profile analyzed
in \cite{Si111bchthy_Liu} if Eq.\ (\ref{indentifyk}) is used to relate
parameters in the two models. As $R\rightarrow 1$,
we have $k\rightarrow \infty $, and we arrive at the local
equilibrium boundary condition with $\bar{c}_{t}^{\pm }=c_{eq}$.

However, if diffusion is {\em faster} in the step region than in the terrace
region ($R<1$), we find a new regime with a {\em negative} kinetic
coefficient. Equivalently, the characteristic length
$d=D_{t}/k$ is negative, but with $d\geq -s/2$. 

The possibility of a negative kinetic coefficient
in the presence of a Schwoebel barrier was
first suggested by Politi and Villain, \cite{negativecoef}
but with no derivation or discussion of any physical consequences.
We argue here that negative kinetic coefficients can play a key role in
understanding current-induced instabilities on Si surfaces.
This has quite different consequences than a model with
permeable steps.\cite{otherattempts} Consider a small perturbation
$\delta x_{n}\left( y,t\right)\equiv x_{n}\left( y,t\right) -x_{n}^{0}=
\varepsilon e^{\omega t+iqy+in\phi}+c.c.$ of the uniform step train.
We report results for a linear stability analysis of the sharp
step model in the physically relevant limits of
weak fields ($fl\ll 1)$ and long wavelengths ($ql\ll 1$). An instability arises
from a positive $\omega =\omega _{1}\left( f,\phi \right) +\omega
_{2}\left( q,f,\phi \right) $, where 
\begin{equation}
\omega _{1}=\Omega D_{t}c_{eq}^{0}\frac{4df}{\left( l+2d\right) ^{2}}\left(
1-\cos \phi \right),  \label{eq:bch_bcf}
\end{equation}
and 
\begin{equation}
\begin{array}{ll}
\omega _{2}= & \Omega D_{t}c_{eq}^{0}q^{2}\left\{ -\Gamma \left[ 
%TCIMACRO{\dfrac{2\left( 1-\cos \phi \right) }{l+2d} }
%BeginExpansion
{\displaystyle {2\left( 1-\cos \phi \right)  \over l+2d}}%
%EndExpansion
+\left( l+%
%TCIMACRO{\dfrac{s}{R} }
%BeginExpansion
{\displaystyle {s \over R}}%
%EndExpansion
\right) q^{2}\right] \right. \\ 
& \left. +f\left[ 
%TCIMACRO{\dfrac{2dl}{l+2d} }
%BeginExpansion
{\displaystyle {2dl \over l+2d}}%
%EndExpansion
+%
%TCIMACRO{\dfrac{s}{R} }
%BeginExpansion
{\displaystyle {s \over R}}%
%EndExpansion
\right] \right\} .
\end{array}
\label{eq:2Dwd_bcf}
\end{equation}

Step bunching is controlled by $\omega _{1}.$ A pairing instability with
maximum amplitude at $\phi =\pi $ is found for $df>0,$ or $(R-1)f>0,$ using
the two-region model parameters. Thus, the profile illustrated in Fig.\ (1),
produced by a step down current and slower diffusion in the step region, is
unstable to step bunching, consistent with the usual interpretation
\cite{Si111bchthy_Liu} of range I in Si(111).
But a similar unstable profile arises from a 
{\em step up current} ($f<0$) along with a {\em negative} $d$ or $k$.

$\omega _{2}$ characterizes 2D step wandering.
The first term in square brackets is always stabilizing
and has its minimum value for in-phase wandering with $\phi =0$.
The next term,
proportional to the field, has two contributions. The first,
proportional to $D_{t}c_{eq}^{0}dfq^{2}$, describes a
Mullins-Sekerka or Bales-Zangwill instability \cite{baleszangwill} induced
by the terrace density field for $df>0$. But as shown above, step
bunching occurs under these same conditions. Wandering of the bunched
steps is generally suppressed, as is seen experimentally in range I of
Si(111).

However, the second contribution, proportional to $D_{s}c_{eq}^{0}sfq^{2}$,
represents an alternate and quite general mechanism for step wandering that is
operative whenever there is a {\em downhill} force ($f>0$).
Downhill step perturbations are amplified by a field-driven downhill flux
of adatoms along the step edge with steady state density per unit step
length approximated by $c_{s}^{0}=c_{eq}^{0}s$.
Using Eq.\ (\ref{indentifyk}) we see that the last term in
Eq.\ (\ref{eq:2Dwd_bcf}) is always positive even when $d<0$ and
the Mullins-Sekerka contribution is stabilizing.

Consider now the implications of these results for Si(111). The low temperature
experiments are well explained by slower diffusion in the step region, 
consistent with the usual picture of an attachment barrier. \cite{Si111bchthy_Liu}
At higher temperatures in range II we suppose that changes in reconstruction
could result in faster diffusion in the step region,
implying a negative $d$ or $k$ in the
sharp-step model. Bunching then is predicted for $f<0$, and step wandering
for $f>0$, in agreement with
experiments \cite{yagireview} and computer simulations \cite{natorisi111sim}
of such a model. The negative kinetic coefficient reverses the bunching
direction, which allows the general wandering instability from a downhill force
to be easily seen. Indeed $d<0$ and
$f>0$ represents the {\em only} case where step wandering occurs with
current in the opposite direction to that giving step bunching. 
Using parameters appropriate for Si(111) we find that the most
unstable wavelength is of order $\mu $m, in
qualitative agreement with experiment.

Since only relative diffusion rates are important, one could imagine
the diffusion rates changing again at higher temperatures so
that $d>0$, possibly describing range III. In this scenario, the transitions
between the different temperature ranges would be associated with local
equilibrium behavior as $R$ passes through unity, where no step bunching or
wandering would be seen.

These ideas also provide an interpretation of electromigration results for
the technologically important Si(001) surface
miscut along $\langle 110\rangle $. At equilibrium rather straight
$S_{A}$ steps that run parallel to the dimer rows of the upper $A$ terrace
alternate with much rougher $S_{B}$ steps that run perpendicular to the
dimer rows of the upper $B$ terrace \cite{krugsi001}. Moreover, diffusion
parallel to the dimer rows is up to a thousand times faster at low
temperatures \cite{yagireview}. For driven diffusion with a current normal
to the steps we thus have $D_{t}^{B}\gg D_{t}^{A},$ and we expect that this
difference will dominate the physics of current-induced instabilities of
Si(001).

To apply our step-region ideas to this case, we imagine as before that a
classical local equilibrium step resides in the center of each step region,
but now let the downhill half-step region differ from the uphill half-step
region, and assume that diffusion in each half-step region is similar to
that in the nearest adjacent terrace.
Defining $R^{i}\equiv D_{t}^{i}/D_{s}^{i}$, with $i=(A,B)$, we can let
$R^{i}$ differ from unity, thus generating asymmetric kinetic
coefficients. We require only that $D_{s}^{B}\geq D_{s}^{A}$, which
seems quite reasonable since $D_{t}^{B}\gg D_{t}^{A}$.
Thus $D_{s}^{B}-D_{s}^{A}=\xi_{s}^{AB}(D_{t}^{B}-D_{t}^{A})$
with $\xi _{s}^{AB}\geq 0$. Special cases of
this model include classical local equilibrium steps where $R^{A}=R^{B}=1$
and a symmetric step model where $D_{s}^{B}=D_{s}^{A}$.

Experiments show that when a direct current is applied to a
configuration of alternating $S_{A}$ and $S_{B}$ steps, the steps move in
opposite directions and step pairs form. With a downhill current one finds
double height $D_{B}$ steps (consisting of an upper $S_{B}$ step and a lower 
$S_{A}$ step with a very narrow $A$ terrace trapped in between) separated by
wide $B$ terraces; the equivalent configuration with $D_{A}$ steps and wide
$A$ terraces is seen for an uphill current. On continued exposure to current,
a step bunching instability of the double height steps
is seen for current in {\em either} direction at low temperatures. \cite
{yagireview,Si001bchexp_Nielsen}

The initial step pairing can be most easily understood by calculating the
velocity of steps in a configuration of equally spaced straight $S_{A}$ and
$S_{B}$ steps. \cite{natorisi001} Using the flux given by Eq.\ (\ref
{steadystateflux}) with the appropriate values of $D_{t}$ and Eq.\ (\ref
{normalvelocity}) we find that the initial velocity of an $S_{B}$ step can be
written as $v_{B}=K_{AB}(D_{t}^{B}-D_{t}^{A})f$, where $K_{AB}$ is
positive and symmetric in $A$ and $B$. The initial velocity of an $S_{A}$
step is given by the same formula when $A$ and $B$ are swapped and hence has
the opposite sign. Thus, for positive $f$, $B$ terraces grow and $A$ terraces
shrink. Given the great difference in $D_{t}^{B}$ and $D_{t}^{A}$, this
process will continue (as is shown by a general analysis with unequal
terrace widths \cite{zwktobepublished})\ until the terraces have very
different sizes, with the final width $l^{\prime }$ of the narrow $A$
terrace in the $D_{B}$ step probably controlled by step repulsions \cite
{natorisi001} (not taken into account in this version of our model).

To explain the continued bunching of the double-height steps, we use the
two-region model, with major terraces separated by a double-height step, which
we treat as a single effective step region. However a minor terrace now
resides in the center of the effective step region, and we must take this
into account in our extrapolation analysis leading to Eq.\ (\ref{indentifyk}).
We can proceed as before if we note that the effective equilibrium
density in the center $\hat{c}_{eq}^{0}$ is linearly modified by the weak
field from its value $c_{eq}^{0}$ at the ``real'' local equilibrium step near
the lower boundary of the effective step region, so that $\hat{c}%
_{eq}^{0}=c_{eq}^{0}[1-f(l^{\prime }+s)/2]$. Thus the analogue of Eq.\ (\ref
{indentifyk}) is \cite{our1dmodel} 
\begin{equation}
\hat{d}^{i}\equiv \frac{D_{t}^{i}}{\hat{k}^{i}}=\frac{1}{2}\left[
R^{i}-\left( 2+\frac{l^{\prime }}{s}\right) \right] s,  \label{keffsi001}
\end{equation}
where $\symbol{94}$ denotes an effective double step parameter.

As shown above, with a step down current ($f>0$) $B$ terraces grow and
$D_{B} $ steps form. According to Eq.\ (\ref{eq:bch_bcf}), step bunching
occurs when $df>0.$ Thus continued bunching of the $D_{B}$ steps requires
that $\hat{d}^{B}$ in Eq.\ (\ref{keffsi001}) is greater than zero, or
$R^{B}>2+l^{\prime }/s.$ (Note that a local equilibrium assumption \cite
{natorisi001} with $R^{B}=1$ can give step pairing, but is inconsistent with
further step bunching.) With step up current ($f<0$) $A$ terraces grow and
$D_{A}$ steps form. Continued bunching now requires that $\hat{d}^{A}$ is 
{\em negative}, or $R^{A}<2+l^{\prime }/s.$ This inequality can be satisfied
even if $R^{A}>1$, and depending on the value of $l^{\prime }$,
could hold under rather general conditions. At higher temperatures we
expect values of $R^{i}$ closer to unity due to thermal fluctuations. If
$R^{B}$ becomes less than $2+l^{\prime }/s$ at some higher temperature, then
Si(001) could exhibit behavior much like range II of Si(111), with bunching
only for a step up current, due to the (effective) negative coefficient for
the $D_{A}$ step. Step wandering of the $D_{B}$ steps from a step down
current, suppressed by the bunching at lower temperatures, would also be
expected.

Generalizations of these ideas and applications
to experiments where the current is directed at an angle to the steps
\cite{Si001bchexp_Nielsen}, along with comparison
to results of Monte Carlo simulations will be presented
elsewhere. \cite{zwktobepublished} We
are grateful to Ted Einstein, Oliver Pierre-Louis, and Ellen Williams
for helpful discussions. This work was
supported by the NSF MRSEC grant DMR 00-80008.


\begin{thebibliography}{10}
\bibitem{our1dmodel}  This result is independent of $f$ and holds under
much more general conditions, as shown by
an exact solution of a 1D model in which adatoms have field-biased
jumps between discrete lattice sites over barriers of different heights in
the terrace and step regions. See \cite{zwktobepublished}.
\end{thebibliography}
\end{document}